\documentclass[12pt,preprint]{aastex}
\newcommand{\PSbox}[3]{\mbox{\rule{0in}{#3}\includegraphics{#1}\hspace{#2}}}
\newcommand{\FigNum}[1]{\unitlength 1pt \begin{picture}(55,10)(-400,35)
                        \put(0,0){Figure #1}
                        \end{picture}}
\newcommand{\perval}[2]{{#1\mbox{$^{#2}$}}}

\newcommand{\persec}{\perval{\rm s}{-1}\/}
\newcommand{\percm}{\mbox{$\cm^{-2}$}}

\newcommand{\ppm}{\mbox{$\pm$}}
\newcommand{\cgsflux}{\erg\,\percm\,\persec}

\newcommand{\cgslum}{\erg~\persec}

\newcommand{\approxlt}{\mbox{$\lesssim$}}
\newcommand{\approxgt}{\mbox{$\gtrsim$}}
\def\etal{{et~al.}}

\newcommand{\nh}{\mbox{$N_{\rm H}$}}
\newcommand{\nhtt}{\mbox{$N_{\rm H, 22}$}}

\def\chisqrnu{\mbox{$\chi^2_\nu$}}
\newcommand{\ud}[2]{\mbox{$^{+ #1}_{- #2}$}}

\newcommand{\ee}[1]{\mbox{$10^{#1}$}}
\newcommand{\tee}[1]{\mbox{$\times 10^{#1}$}}

\newcommand{\keV}{\mbox{$\rm\,keV$}}

\newcommand{\cm}{\mbox{$\rm\,cm$}}

\newcommand{\erg}{\mbox{$\rm\,erg$}\/}

\newcommand{\mJy}{\mbox{$\rm\,\mu Jy$}}

\newcommand{\chandra}{{\em Chandra\/}}
\newcommand{\rosat}{{\em ROSAT\/}}

\newcommand{\rxte}{{\em RXTE\/}}
\newcommand{\xmm}{{\em XMM\/}}

\newcommand{\swift}{{\em SWIFT\/}}
\newcommand{\kpc}{\mbox{$\rm\,kpc$}}

\newcommand\musec{$\mu$s}
\newcommand\msec{{ms}}

\newcommand{\fxfr}{\mbox{$F_{X}/F_{\rm 1.4\,GHz}$}}
\def\cxo{CXOU J191121.3+003844}  
\def\xray{X-ray}
\def\rrat{RRAT~J1911+00}
\def\aql{Aql~X-1}
\received{}
\revised{}
\slugcomment{}

\begin{document}

\title{Observational Limits on X-ray Bursts from \rrat}

\author{Kelsey Hoffman, Robert E. Rutledge\altaffilmark{1}, 
Derek B. Fox\altaffilmark{2}, 
Avishay Gal-Yam\altaffilmark{3}, 
S. Bradley Cenko\altaffilmark{4}
}

\altaffiltext{1}{Department of Physics, McGill University, 3600 rue
  University, Montreal, QC, H3A 2T8, Canada; hoffmank, rutledge (@physics.mcgill.ca)}
\altaffiltext{2}{Department of Astronomy and Astrophysics, 525 Davey
  Laboratory, Pennsylvania State University, University Park,
  Pennsylvania 16802, USA}
\altaffiltext{3}{Hubble Fellow, Astronomy Department, MS 105-24, California Institute
  of Technology, Pasadena, CA 91125, USA}
\altaffiltext{4}{Space Radiation Laboratory, MS 220-47,
  California Institute of Technology, Pasadena, CA 91125}

\begin{abstract}
The high radio-flux brightness temperature of the recently discovered
class of sources known as Rotating RAdio Transients (RRATs) motivates
detailed study in the X-ray band.  The source \rrat\ has a (large)
error region which includes the frequently observed low-mass X-ray
binary transient \aql.  We describe analyses of historical X-ray data,
searching for X-ray phenomena (sources, behaviors), finding no sources
or behaviors which may unequivocally be associated with \rrat.  We put
forward a candidate X-ray counterpart to \rrat, discovered in a
\chandra\ observation in Feb 2001, which fades by a factor $>5$ prior
to April 2004. Archival {\em ROSAT} observations detect the source as
early as Oct 1992, with a flux comparable to the Feb 2001 {\em
Chandra} detection. The X-ray flux and optical ($F_X/F_R>12$) and near
infra-red ($F_X/F_J>35$) limits, as well as the X-ray flux itself, are
consistent with an AGN origin, unrelated to \rrat.  Searches for msec
X-ray bursts found no evidence for such a signal, and we place the
first observational upper-limit on the X-ray to radio flux ratio of
RRAT bursts: $\fxfr<6 \tee{-11} \cgsflux\ {\rm mJy}^{-1}$.  The
upper-limit on the X-ray burst flux (corresponding to $<2.2\tee{37}
(d/3.3\,\kpc)^2 \cgslum$, 2-10 keV) requires a limit on the spectral
energy density power-law slope of $\alpha<-0.3$ between the radio and
X-ray bands. We place a limit on the time-average X-ray burst
luminosity, associated with radio bursts, of
$\leq3.4\tee{30}\,(d/3.3\,\kpc)^2 \cgslum$. Future definitive
association between \cxo\ and \rrat\ could be made with a better radio
localization of \rrat; or discovery of msec X-ray bursts associated
with the radio bursts.  Adaptive optic infra-red observations of the
crowded source field of \cxo\ will be required to find a NIR
counterpart, which may reveal its source class.
\end{abstract}

\keywords{stars: flare; stars: neutron; X-rays: stars; pulsars:
  individual (J1911+00)}

\section{Introduction}

Recently, a new observational class of neutron stars has been proposed
\citep{mclaughlin06}, comprising eleven sources defined by the
phenomenon of fast (2-30 ms) radio bursts, which repeat on timescales
of 4\,min - 3\,hr. Through period folding, periodicities were
claimed to be found in the range of $P=$0.4-7\,s for ten of the
eleven sources, which suggested association with rotating neutron
stars.  In three of the eleven sources, period derivatives were
claimed, with one suggesting a magnetic field strength of 5\tee{13} G
-- a possible magnetar.  Because the $P-\dot{P}$ analysis remains
unpublished and undescribed, we regard association of the
radio-bursting population with neutron stars to be an unsupported
hypothesis; nonetheless, we adopt the hypothesis and nomenclature of
referring to the sources as Rotating RAdio Transients (RRATs), and
consider their association with neutron stars to be the leading
hypothesis.

The proposed class was recently made more intriguing by the discovery
of transient pulsed radio emission from a known magnetar,
XTE~J1810-197 \citep{camilo06}, with peak flux density $>$1 Jy and
89\ppm5\% polarization at 8.4 GHz, with period $P\sim5.54\, {\rm s}$.
The radio pulsation was not present prior to the 2003 discovery
outburst of this magnetar \citep{ibrahim04}, detected serendipitously
from X-ray pulsations during observations of a nearby source;
sensitive X-ray observations found a fading two-temperature X-ray
source, with luminosity $\approx\ee{34-35}$ \cgslum\ on timescales of
300 and 900 days (for two different temperature components;
\citealt{gotthelf05}).  Moreover, day-to-day fluctuations in the
radio peak flux were observed, of $\approx$2. This is interpreted to
imply the appearance of the radio pulsations is associated with the
reconfiguration of the magnetar magnetic field, which takes place
associated with magnetar bursting activity \citep{woods01}.  The
relationship between the observed process and RRATs is unclear,
however observation of radio and X-ray activity associated with
transient magnetic field reconfiguration may be applicable to the
RRATs, in light of the observed transient nature of their radio emission. 

An X-ray counterpart has been reported for RRAT J1819-1458, discovered
serendipitously as CXOU J181934.1$-$145804 \citep{reynolds06}.  This
X-ray source was observed with absorption corrected flux 2\tee{-12}
\cgsflux\ (0.5-8 keV, with an order of magnitude uncertainty due to
uncertainty in the absorption) detected within the 5\arcsec\ $\times$
32\arcsec error ellipse of the RRAT.  Comparison with the source
density of such X-ray sources find the probability of finding such a
bright source by chance $<$\ee{-4}, compared with the {\em ASCA}
Galactic Plane Survey \citep{sugizaki01}, at the best-fit unabsorbed
flux of 2\tee{-12} \cgsflux.  The X-ray spectrum is soft
($kT$=120\ppm40 eV), which supports classification of a thermal,
cooling isolated neutron star (INS) -- further supported by the
absence of a positionally coincident optical or infrared counterpart.
The most convincing evidence for association CXOU J181934.1$-$145804
with RRAT J1819-1458 is the low probability of finding such a bright
X-ray source in the RRAT positional error circle.  As Reynolds \etal\
point out, the observed soft spectrum in light of the high Galactic
column density (\nh=1.6\tee{22} \perval{cm}{-2}) in the direction of
this source strongly supports interpretation of the X-ray source as
within the Galaxy, further supporting an neutron star interpretation.

A second source, \rrat, has an error ellipse of
15\arcmin$\times$7\arcmin (1$\sigma$), which includes the prolific
transient neutron star low-mass X-ray binary Aql X-1 \citep{liu01}.
The derived dispersion measure distance \citep{mclaughlin06} is
$d=$3.3 kpc, modestly but perhaps not significantly closer than the
typical 4-6 kpc distance adopted for Aql X-1.  Four radio bursts at
$\nu=$1.4 GHz, of half-width $w_{50}$=2 msec were observed in 13 hours
of observing -- the least active of the tabulated RRATs -- with peak
fluxes of $S_\nu$=250 mJy.  If we take $w_{50}$ to be the light
crossing time of the emission region, this implies a source brightness
temperature (e.g., \citealt{guedel02}):

\begin{equation}
T_b = 4\times10^{22} \left(\frac{S_\nu}{250\, {\rm mJy}}\right) 
\left( \frac{ 1.4\, {\rm GHz}}{\nu}\right)^2
\left( \frac{2\,{\rm ms}}{w_{\rm 50}} \right)^2
\left( \frac{d}{3.3\, {\rm kpc}} \right)^2 \, {\rm K }
\end{equation}

\noindent  This is significantly above the brightness temperature
limit for inverse Compton scattering ($T_B\sim\ee{12} K$)
which likely requires a coherent emission mechanism (for this, and
other RRATs), possibly accompanied by X-ray emission.  The historical
X-ray coverage of Aql X-1 permits study of this field (although not
covering the entire error box of the RRAT).

In this paper, we report X-ray studies of the field of \rrat.  We
describe a candidate X-ray counterpart (\cxo) to the \rrat, detected
serendipitously in an historical {\em Chandra} observation of Aql X-1
(\S~\ref{sec:obs}).  In \S~\ref{sec:pca}, we describe a search for
msec X-ray bursts from the RRAT, deriving the first limit on the X-ray
to radio flux ratio for the RRAT bursts.  In \S~\ref{sec:con} we
summarize and discuss our results.

\section{Observations and Analysis}
\label{sec:obs}

For analyses, we used CIAO
v3.3\footnote{http://cxc.harvard.edu/ciao/},
FTOOLS\footnote{http://heasarc.gsfc.nasa.gov/ftools/} \citep{ftools},
and XSPEC v11.3.1 \citep{xspec}, with common tasks as described.  We
found existing observations from \chandra, \xmm, \rosat, {\em SWIFT},
\rxte, and {\em Beppo}SAX in the public archive at
HEASARC\footnote{http://heasarc.gsfc.nasa.gov/}
(Table~\ref{tab:observations}).  While there were 11 \chandra\ ACIS-S
imaging observations in the archive, we tabulate only the one
observation in which a second X-ray source (that is, other than \aql)
was found, as described in the following section.

\subsection{Chandra Observation}

We used the CIAO tool {\tt celldetect} with event level=1 and event
level=2 observational data to search for sources in the 11 existing
ACIS-S \chandra\ observations, taken between Nov 2000 and Sept
2002. The $1/2\arcmin\times4\arcmin$ FOV of these observations rotates
on the sky so that, except for an area $\sim
1/2\arcmin\times1/2\arcmin$ near \aql, the observations cover
different areas of the sky.  The event level=1 data was used for the
source search, in addition to the processed event level=2 data, since
msec duration X-ray bursting activity could result in pile-up; events
associated with a candidate RRAT counterpart may be flagged as cosmic
ray events and filtered from the observation data.  With a
signal-to-noise limit of $>4$, we found no sources which appeared in
level=1 and not level=2 data in any of 11 \chandra/ACIS-S imaging
observations.  We also found two sources which appeared in both the
level=1 and level=2 data, one of which is \aql.  The second X-ray
source is localized R.A.=19h11m21s.36, dec.=00d38m43.69s (J2000)
\ppm0.6\arcsec\ (90\% confidence), approximately 3.9\arcmin\ from Aql
X-1.  The position of this object was not within the FOV of any of the
other 10 existing \chandra\ observations.  No USNO-B1.0 or 2MASS
source is cataloged within 1\arcsec\ of this position. We designate
the unidentified X-ray source as \cxo, and performed a more detailed
analysis to search for RRAT or neutron-star type properties.

\subsubsection{\cxo}
We extracted data within 4\arcsec\ of \cxo, finding 100 counts in the
source region; we neglect background counts ($\sim$3 counts)
throughout the analysis.

{\bf Spectral Analysis.}  We performed spectral analysis, extracting a
spectrum of \cxo\ from the event level=2 data, using data in the
0.5-10 keV range, and grouping PI bins to have between 18 and 21
counts and an energy width greater than the energy resolution at the
average photon energy.  A systematic uncertainty of 4\% was applied to
the spectrum and the absorption was held fixed at \nh=3.3\tee{21}
\perval{cm}{-2} \citep{dickey90}.  Results are shown in
Table~\ref{tab:cxospec}.  An absorbed blackbody model does not produce
an acceptable fit (\chisqrnu=4.47, 3 degrees of freedom, or dof,
Prob=$4\times10^{-3}$). A power-law model fits acceptably
(\chisqrnu=0.65, 3 dof, Prob=0.59) with a photon power-law slope
$\alpha$=1.0\ud{0.3}{0.4}, and an unabsorbed flux of 1.8\tee{-13}
\cgsflux (0.5-8 keV), for a luminosity of 2.4\tee{32} $(D/3.3\, {\rm
kpc})^2$ \cgslum.

{\bf Pulsation Search}.  To search for pulsations, we produced a power
density spectrum (PDS; \citealt{press}) using FFTW \citep{fftw}, with
frequency resolution of 0.116 mHz, and a Nyquist frequency of 1.13368 Hz.
We find no evidence of coherent pulsations in the 0.0011-1.133 Hz
range, with an upper-limit (based on the greatest observed power) of
$<$42\% rms.  The observation is not very sensitive due to the low number
of counts.

{\bf Bursting Activity}. Due to the $\sim0.4\,{\rm s}$ time resolution
of the \chandra\ data, a $\sim$msec burst of (multiple) X-rays, such
as might accompany the observed radio bursts, could result in
instrumental pile-up -- in which multiple X-ray photons are counted as
a single photon, their photo-electrons summed into a single count,
which is assigned a commensurately higher photon energy. Such a count
could be flagged as a cosmic ray and removed between the level=1 and 2
data processing; or it could be assigned a to high Pulse Invariant
(PI) channels in the level=1 data and remain in the level=2 data.  To
investigate $\sim$msec bursting activity of \cxo, with this data the
high PI channels of the event level=1 observation were compared to the
background.

In a 4\arcsec\ region about the source, event level=1 data contained
109 counts, while the event level=2 contained 100 counts, for 9 counts
rejected as cosmic rays. In a 33\arcsec\ region away from the source,
event level=1 data contained 860 counts, while event level=2 data
contained 247 counts; this corresponds to an average 9\ppm0.3 events
rejected between level=1 and level=2 data in our source region --
consistent with that observed.  We therefore find no evidence of msec
X-ray bursts (rejected as cosmic rays) in the source region.

{\bf Optical/NIR Flux Ratios.}  We obtained $J$-band imaging of \cxo\
using the Wide-field Infrared Camera (WIRC; \citealt{wirc}) at prime
focus of the Palomar Hale 200-inch telescope on 20~July 2006 UT (mean
epoch 08:30 UT).  We took individual exposures of one minute (as
$2\times 30$\,s), obtaining 20 dithered exposures in all.  Individual
frames were dark-subtracted and flat-fielded using afternoon
calibrations; a fringe image was created from sky exposures and
subtracted from the frames; bad pixels and cosmic rays in individual
frames were identified and masked, and the frames registered and
combined.  Final astrometric and photometric calibration was performed
against the 2MASS catalog.  Since \aql\ is detected in the final WIRC
image, we are able to register directly against the \chandra\ image,
localizing \cxo\ to within $\pm$0.3\arcsec; we find a position for the
source of R.A. 19:11:21.41, Dec +00:38:44.5 (J2000).  In
Fig.~\ref{fig:opt}(a) we present a portion of the WIRC image including
the position of \cxo.  We observe no NIR counterpart to $J>21.2$\,mag.

We then obtained $r'$-band imaging of the field using the Large Format
Camera
(LFC\footnote{http://www.astro.caltech.edu/palomar/200inch/lfc/index.html})
at prime focus of the Hale Telescope on 1~August 2006 UT (mean epoch:
08:34 UT).  We obtained $6\times 120$\,s dithered exposures which were
then bias-subtracted, flat-fielded, registered against the NOMAD
catalog, and combined (with bad pixel and cosmic ray masking).  A
portion of the resulting image is shown in Fig.~\ref{fig:opt}(b).
Although scattered light from nearby stars is visible at the position
of \cxo, no point-like source is apparent.  Photometric registration
against the NOMAD catalog (with $R$-band magnitudes from USNO-B1.0)
allows us to set a limit of $R>22.8$\,mag on the brightness of any
point source.

To calculate \xray\ to optical/NIR flux ratios for \cxo, it is
necessary to correct our observed limits for Galactic extinction.
Since the source lies in the plane of the Galaxy, the total Galactic
reddening is uncertain; however, the estimate from \citet{schlegel98}
is $E(B-V)=0.73$\,mag, which implies extinctions of $A_R=1.95$\,mag
and $A_J=0.66$\,mag, respectively.  Applying these gives
extinction-corrected limits of $R>20.8$\,mag and $J>20.5$\,mag for any
point-source counterpart to \cxo.  Constructing the ratio from the
source's observed 2--10~keV \xray\ flux to our extinction-corrected
limits, we then find $F_{\rm X}/F_R > 12$ and $F_{\rm X}/F_J > 35$;
these values are inconsistent with those for clusters, stars, and
isolated white dwarfs (see, for example, \citealt{beuermann99};
\footnote{compare with http://heasarc.gsfc.nasa.gov/wgacat/about.html,
Figure 13}), but are consistent with QSOs/AGN, CVs and other X-ray
binaries. 

\subsubsection{\aql}

As Aql X-1 also fell within the error ellipse of \rrat, activity from
Aql X-1 that could be associated with a $\sim$ msec radio burst was
investigated.  There are 11 historical \chandra\ observations of Aql
X-1 taken with ACIS-S. \cxo\ is only in one of these observations and
the coordinates of \cxo\ do not fall within in the FOV of any of the
other 10 \chandra\ observations. Using these 11 \chandra\
observations, source counts were extracted from the event level=1
files using a 5\arcsec\ radius extraction region, centered on Aql
X-1. Background counts from a circle of 33\arcsec\ far from the source
were also extracted from each of the 11 observations. All the counts
in each PI channel were summed from each of the observations for the
source and the background. Then the PI channels were binned such that
16 channels became one bin. In the highest binned PI channel there are
a total of 48 counts from the Aql X-1 regions (source plus background)
and 38 background region area corrected counts. Treating the
background counts as the mean, the probability of getting 48 counts or
more is 7$\%$ using the Poisson probability distribution. Thus, no
significant counts in the high PI channel are found in excess of that
expected from the background.

\section{Historical X-ray Flux of \cxo}

We examined the history of X-ray observations of this source.  These
are described in Table~\ref{tab:observations}, including the dates of
observations, instruments used, the observation duration, the
energy-counts ratio (ECR) for converting countrate to flux, and the
corresponding flux (or upper limit) for \cxo.  Below, we give
pertinent details for the individual observations, sorted by
observatory.

We did not analyse data from {\em R}XTE for a flux limit, as we found
the observed flux from the candidate counterpart produce a countrate
($\approx$0.01 c/s), well below the detection limit, even for pulsed
emission, in $<$\ee{5} sec; the required integration times are even
longer when Aql X-1, which shares the {\em R}XTE FOV, is active.

We examined visually but did not anaylse in detail four {\em Beppo}SAX
observations taken during March 8-20 1997, associated with an
outburst of \aql.  The large point spread function (9.7\arcmin\ FWHM in
LECS; 75\arcsec\ half-power radius in MECS) overlapping PSF of Aql X-1
(3.9\arcmin\ away) at the source position precludes source detection
at fluxes comparable to that observed in other observations.  However,
visual inspection of LECS and MECS images shows no evidence for any
X-ray source of flux comparable to that of Aql X-1 during any
observing epoch.

{\bf ROSAT}.  For the PSPC observations, we used a source extraction radius of
15\arcsec\ about the position of the candidate, and background from an
annulus 20\arcsec and 75\arcsec\ inner- and outer-radius centered at
the source position.  

{\bf XMM}.  \cxo \ was not detected in either of the two existing \xmm\
observations. Background counts were extracted from an annulus of
20\arcsec\ and 40\arcsec\ inner- and outer-radius (respectively)
centered on \cxo.  

{\bf SWIFT}.  We used an extraction radius of 30\arcsec\ about the
source position, and background from an annulus of inner- and
outer-radius of 40\arcsec and 150\arcsec, respectively.  Of two
observations, separated by 7 days in 2006, no X-ray source is detected
at the position of the \chandra\ source.

The resulting X-ray measured X-ray fluxes and flux limits are given in
Table~\ref{tab:observations}, and are shown in Fig.~\ref{fig:flux}.
  The X-ray source was detected as early as Oct 1992, and as recently
  as Feb 2001.  However, 90\% confidence limits measured with \xmm\
  and \swift\ show the X-ray source has faded since its Feb 2001
  detection using \chandra; comparison with the results of the 2004
  Apr 19 \xmm\ observation show that the source faded by a factor of
  \approxgt 5. 

\section{\rxte/PCA X-ray Burst Search}
\label{sec:pca}
We also searched for X-ray bursts in \rxte/PCA data while the
instrument is pointed at Aql X-1.  We examined 25.6 hours of data from
39 observations in the public
Archive\footnote{http://legacy.gsfc.nasa.gov}, taken with the
Proportional Counter Array \citep{pca} and data acquired in Event mode
{\tt E\_125\_64M\_0\_1\_s, E\_500us\_64M\_0\_1\_s}, using PHA channels
5-27 only; these correspond, depending on the observational epoch, to
photon energy ranges of 1.5-7.4 keV, up to 2.0-11.5 keV.  We excluded
event channel 0 (corresponding to PHA channels 0-4), as there are
frequent burst-like-events in this channel which appear in
observations of other objects as well; and which are largely
concentrated in the Proportional Counter Units (PCUs) PCU1, PCU3 and
PCU4. These events are understood to be due to high-voltage breakdown
in these detector units.  We selected event channel 16 (PHA channel
27) as the upper energy channel, to minimize background while
permitting observational bandwidth over an astrophysically interesting
range.

Our search method proceeded as follows: we used data from 3, 4, or 5
PCUs, the number of which depended on how many were in operation
during the selected observation.  We examined observations in which
the average countrate was $\leq 400$ c/s following energy selection, a
compromise between burst-flux sensitivity and high integration time.
For each count, we measured the associated average countrate from
within \ppm1 second, producing a 2-second running average countrate.

\newcommand{\ptrig}{\mbox{$P_{\rm Trigger}$}} At the time of each
count, we found the number $N$ all counts within $\ppm1.106\,{\rm
ms}$, excluding the initial count (that is, there are $N+1$ counts in
each period). We calculated the probability \ptrig\ of observing $N$
counts from a non-variable countrate, where we assumed the average
countrate is the associated average countrate within $\pm1\,{\rm s}$
of the count.  When $\ptrig\times N_{\rm phot}<1$, where $N_{\rm
Nphot}$ is the number of counts in the particular individual
observation, we made note of the event as a possible trigger to be
examined later.

Finally, we filtered out triggers in which counts are found improbably
concentrated in a single PCU detector -- presumably, due to a detector
high-voltage breakdown.  When one of the $N_{\rm PCU}$ PCUs contains
the most counts $k$, we find the probability $P$ for $\geq k$ of $N$
counts to be found in a single PCU:

\begin{eqnarray}
P(N, N_{\rm PCU}, k)  = & P_T(N, N_{\rm PCU}, k)/P_T(N, N_{\rm PCU}, 0) &\\
P_T(N, N_{\rm PCU}, k)  = & \sum_{i=k}^{i=N} B(N, N_{\rm PCU}, i)\times B_2(N-i,
N_{\rm PCU}-1, i) & \\
B(N, N_{\rm PCU}, k)  = & \frac{N!}{k!(N-k)!} \left(\frac{1}{N_{\rm
PCU}}\right)^k \left(1 - \frac{1}{N_{\rm PCU}}\right)^{N-k} & \\
B_2(N, N_{\rm PCU}, k)  = &\sum_{i=0}^{i=k} B(N, N_{\rm PCU}, i)
B_2(N-i, N_{\rm PCU}-1, i) &, N_{\rm PCU} \neq 1 \\
                         = &  1 &, N_{\rm PCU}=1, k=N \\
			 = & 0 &, N_{\rm PCU}=1, k\neq N 
\end{eqnarray}

\noindent . Here, $P_T(N, N_{\rm PCU}, k)$ is the un-normalized
probability of observing $\geq k$ counts (of $N$ total counts) in a
single PCU, when the $N$ counts are distributed randomly among the
$N_{\rm PCU}$ PCUs; $B(N, N_{\rm PCU}, k)$ is the binomial probability
of observing $k$ counts (of $N$ total counts) in one particular PCU
(of $N_{\rm PCU}$ PCUs); and $B_2(N, N_{\rm PCU}, k)$ is the
probability of observing {\em no more} than $k$ counts in a particular
PCU nor in the remaining $N_{\rm PCU}-1$ PCUs.  When $P(N, N_{\rm
PCU}, k)\leq0.01$, we considered the counts to be improbably
concentrated to a single PCU to be a celestial X-ray burst (in which
the counts should be distributed evenly among the PCUs, on average),
and the trigger was filtered out as likely due to a detector-related
effect.  We expect only 1\% of all celestial X-ray bursts would match
this trigger, and so it would negligibly affect our detected burst
rate.

In the 25.6 hr of data, we found 6 possible triggers; the most
significant of these had a chance probability of being observed from a
Poisson fluctuation during the single observation of 4\% -- consistent
with a Poisson fluctuation expected from examining 39 observations.
We thus find no evidence of celestial X-ray bursts on a 2 msec
timescale in this field.

We place a limit on the burst rate as a function of burst flux which
could have passed undetected during these observations.  We would have
considered a burst to be detected if the number of counts in a burst
event would only have been produced across the entire dataset from a
background fluctuation in a dataset 100$\times$ as large, on average.
Since zero bursts met this criteria, we take the 95\% confidence
upper-limit to the burst rate to be $2/T(F)$, where $T(F)$ is the
total integrated time as a function of average burst flux $F$ (there
is a 5\% probability of observing zero events when 2 are expected). We
find the 2-10 keV flux $F$ from the count rate limits derived from the
observation, $F= (I/(1000 {\rm c/s/PCU}) \ee{-8} \cgsflux$, where $I$
is the burst countrate limit (averaged over 2.2msec); we assumed a
photon power-law spectral slope of $\alpha=1$ and \nhtt=0.3.

During 25.6 hr of observation, we place a 90\% confidence upper-limit
on the flux of any 2msec burst which occurred as $<$1.7\tee{-8}
\cgsflux\ ($<$0.9 mJy in the 2-10 keV region), or $<2.2\tee{37}
(d/3.3\,\kpc)^2 \cgslum$. Assuming a flux density in power-law form
($F(\nu)=A\nu^{\alpha}$ \mJy) and comparing this limit at $2\,\keV$,
with the radio detection at $1.4\,{\rm GHz}$, we derive an upper-limit
of $\alpha<-0.3$. For a burst rate comparable to that observed from
\rrat\ (0.1 hr$^{-1}$), the minimum detectable burst flux is $\approx
1.5\tee{-8} \cgsflux$, 2-10 keV).  If we assume a radio burst rate of
0.08 hr$^{-1}$ for a radio flux of 250 mJy at 1.4 GHz (comparable to
that observed from \rrat), then we expect that 2 such bursts, on
average, would have gone off during our observations, and the
probability that zero such bursts would have gone off is $<5\%$.  We
therefore can place an upper-limit on the X-ray/radio flux ratio of
the RRAT bursts:

$$
\fxfr \approxlt\ 6\tee{-11} \cgsflux\,{\rm mJy}^{-1}
$$

This corresponds to \fxfr$\approxlt10^{15.8}$ Hz. This limit is
comparable to the measured \fxfr\ for stellar coronal flares (which
are phenomenologically unlike the radio bursts of RRATs, in that their
durations are typically \ee{2-3} sec) \citep{guedel93}.  

This \fxfr\ upper-limit is significantly above the peak pulsed value
for millisecond pulsars (MSPs).  For example, PSR~B1821-24 has a peak
countrate, observed with \chandra/HRC-S, of 62 counts in the
30.54\musec\ peak bin (1\% duty cycle) of the 52586 sec observation,
(0.117 c/s), for 3.5\tee{-12} \cgsflux (2-10 keV), with 13\%
uncertainty, ignoring spectral uncertainty \cite{rutledge04}. 
The peak radio flux (1.4 GHz) is $\sim$70 mJy, with 15\% uncertainty
\cite{cognard96}; this corresponds to $\fxfr = 5\tee{-14} \cgsflux
\perval{\rm mJy}{-1}$ -- a factor of 1200 below than the limit we set
here for \rrat. 

While magnetars would be a useful source class to compare with this
X-ray/radio flux ratio --- in particular, the $\msec$ bursts observed
from anomalous X-ray pulsars (AXPs;
\citealt{gavriil02,kaspi03,gavriil04,woods05}) or those from
soft-gamma-ray repeaters (SGRs; \citealt{gogus99,gogus01}) -- no
published limits on the X-ray/radio ratio of the msec bursts from
magnetars exist.

With a measured limit for the X-ray burst flux of $F_{\rm
lim}<$1.5\tee{-8} \cgsflux (2-10\,\keV); burst duration of 2\,\msec ,
and an average burst rate of 4/13\,\perval{hr}{-1}, this implies an
upper-limit to the time-averaged X-ray burst luminosity of $\leq
3.4\tee{30} (d/3.3\kpc)^2\,\cgslum$.

\section{Discussion and Conclusions}
\label{sec:con}

The historical X-ray coverage of the error box of \rrat\ exists due to
the source's close proximity to Aql X-1, a transient LMXB.  Other than
Aql X-1 itself, we find one unidentified X-ray source in the {\em
Chandra} observations, which has the following properties: moderately
hard spectrum ($\alpha$=1 photon power-law index); intensity
variability by a factor of $\approxgt$5 over timescales of \approxlt
yrs.  We find no evidence of pulsations, but with weak limits
(\approxlt 42\% rms) in the 0.0011-1.133 Hz range.

The observed X-ray flux gives a value of $F_X/F_R>12$, and $F_X/F_J>
35$, limits which are consistent with an AGN, as well as with the much
higher such ratios of clusters, CV/low-mass-X-ray binaries, and
isolated neutron stars.  Based on the density of AGN, the probability
of finding one such object in the $15\arcmin\times7\arcmin$ error box
of \rrat\ is $\sim$0.3 \citep{hasinger01}, which is sufficiently large
as to permit an explanation for this source as simply an unrelated AGN
in the source field.  We thus do not rule out an AGN origin for this
X-ray source   Though we regard the suggestion that RRATs are neutron
stars to be the leading hypothesis, we would not exclude the
possibility that one of the radio-bursting sources could be of a
different origin.

Based on the fading X-ray behavior, we put forward \cxo\  as a
candidate X-ray counterpart to \rrat, warranting further study. Both
\cxo\ and Aql X-1 fall within the error ellipse of \rrat. No new
activity was detected from Aql X-1.  Observations which would securely
identify this object as the counterpart include a 1\arcsec\ radio
localization of \rrat, securing positional coincidence; or
detection of fast ($\sim$\msec) X-ray bursts, which would be associated
with the radio-burst phenomena.  

The limit we place on the X-ray/radio flux ratio of the RRAT bursts,
while the first such limit for this phenomenon, does not seem to
exclude any known transient phenomena; the limit can be used to plan
future, more sensitive searches.

\acknowledgements

RER acknowledges support from NSERC.  AG acknowledges support by NASA
through Hubble Fellowship grant \#HST-HF-01158.01-A awarded by STSci,
which is operated by AURA, Inc., for NASA, under contract NAS
5-26555. SBC is supported by a NASA Graduate Student Research
Fellowship.  The authors acknowledge and thank Caltech Astronomy for
their generous and sustained policy of time allocations for
postdoctoral researchers.  This research has made use of the NASA/
IPAC Infrared Science Archive, which is operated by the Jet Propulsion
Laboratory, California Institute of Technology, under contract with
the National Aeronautics and Space Administration.  This publication
makes use of data products from the Two Micron All Sky Survey, which
is a joint project of the University of Massachusetts and the Infrared
Processing and Analysis Center/California Institute of Technology,
funded by the National Aeronautics and Space Administration and the
National Science Foundation.  The Digitized Sky Surveys were produced
at the Space Telescope Science Institute under U.S. Government grant
NAG W-2166. The images of these surveys are based on photographic data
obtained using the Oschin Schmidt Telescope on Palomar Mountain and
the UK Schmidt Telescope. The plates were processed into the present
compressed digital form with the permission of these institutions.
The DPOSS project was generously supported by the Norris Foundation.

\clearpage

\begin{figure}[htb]
\caption{ \label{fig:flux} X-ray flux history of \cxo.  Upper-limits
  are 90\% confidence. These values are also given in
  Table~\ref{tab:observations}. 
}
\end{figure}

\begin{figure}[htb]
\caption{ \label{fig:limits} 95\% confidence upper-limit burst rates
  for 2-10 keV X-ray bursts (2 ms duration) for an on-axis source
  during 25.6 hrs of \rxte/PCA observations of Aql X-1.  The field of
  view (FOV) has 50\% response at 30\arcmin, and therefore includes
  the position of \rrat.  }
\end{figure}

\begin{figure}[htb]
\caption{ \label{fig:opt} Optical and near-infrared images of the region of sky
  surrounding \cxo.  (a) Near-infrared ($J$-band) image from our Hale
  Telescope + WIRC observations.  The seeing in this image is
  0.9\arcsec; no counterpart is identified to $J>21.2$\,mag as
  calibrated against the 2MASS catalog.  (b) Optical ($r'$-band) image
  from our Hale Telescope + LFC observations.  The seeing in this
  image is 1.3\arcsec.  No counterpart is identified to $R>22.8$\,mag
  as calibrated against the \mbox{USNO-B1.0} catalog.  Note that
  although scattered light from nearby bright stars is visible at the
  position of \cxo, no point-like object is seen.  The total Galactic
  reddening in this direction is uncertain but estimated to be
  $E(B-V)=0.73$\,mag, yielding extinction estimates of $A_R=1.95$\,mag
  and $A_J=0.66$\,mag, respectively \citep{schlegel98}.
}
\end{figure}

\clearpage
\pagestyle{empty}
\begin{figure}[htb]
\PSbox{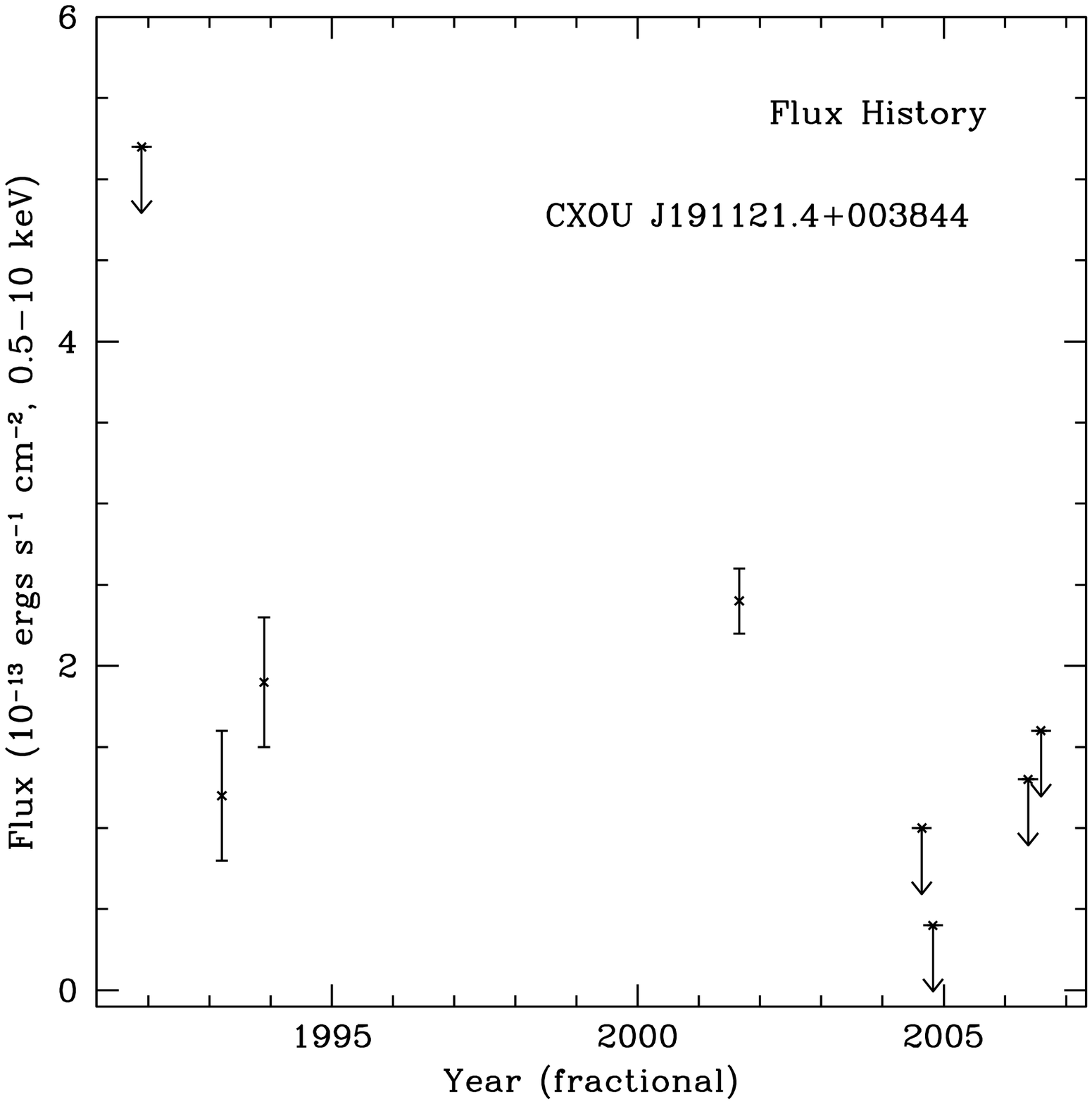 hoffset=-80 voffset=-80}{14.7cm}{21.5cm}
\FigNum{\ref{fig:flux}}
\end{figure}

\clearpage

\clearpage
\pagestyle{empty}
\begin{figure}[htb]
\PSbox{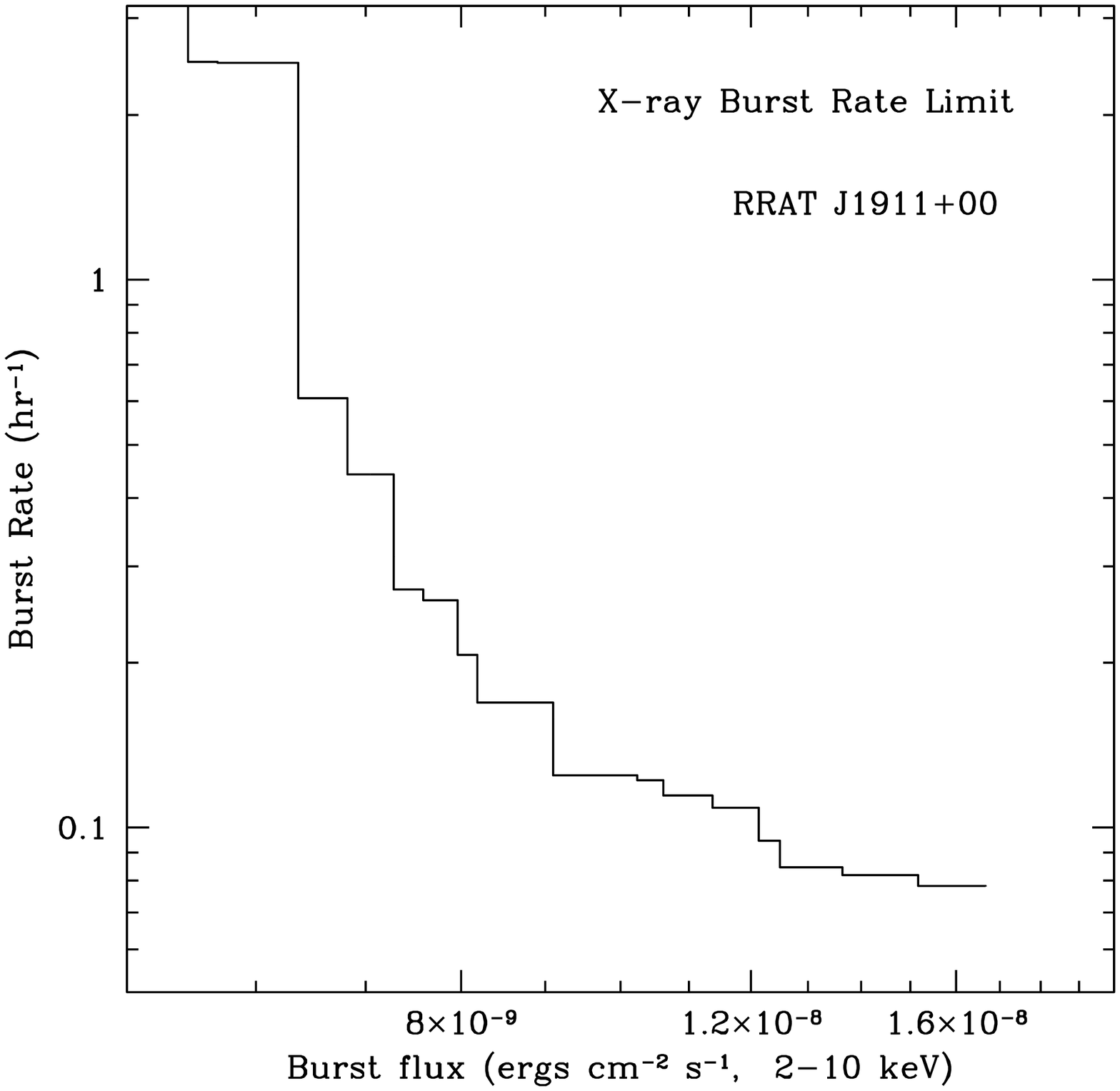 hoffset=-80 voffset=-80}{14.7cm}{21.5cm}
\FigNum{\ref{fig:limits}}
\end{figure}

\clearpage

\clearpage
\pagestyle{empty}
\begin{figure}[htb]
\PSbox{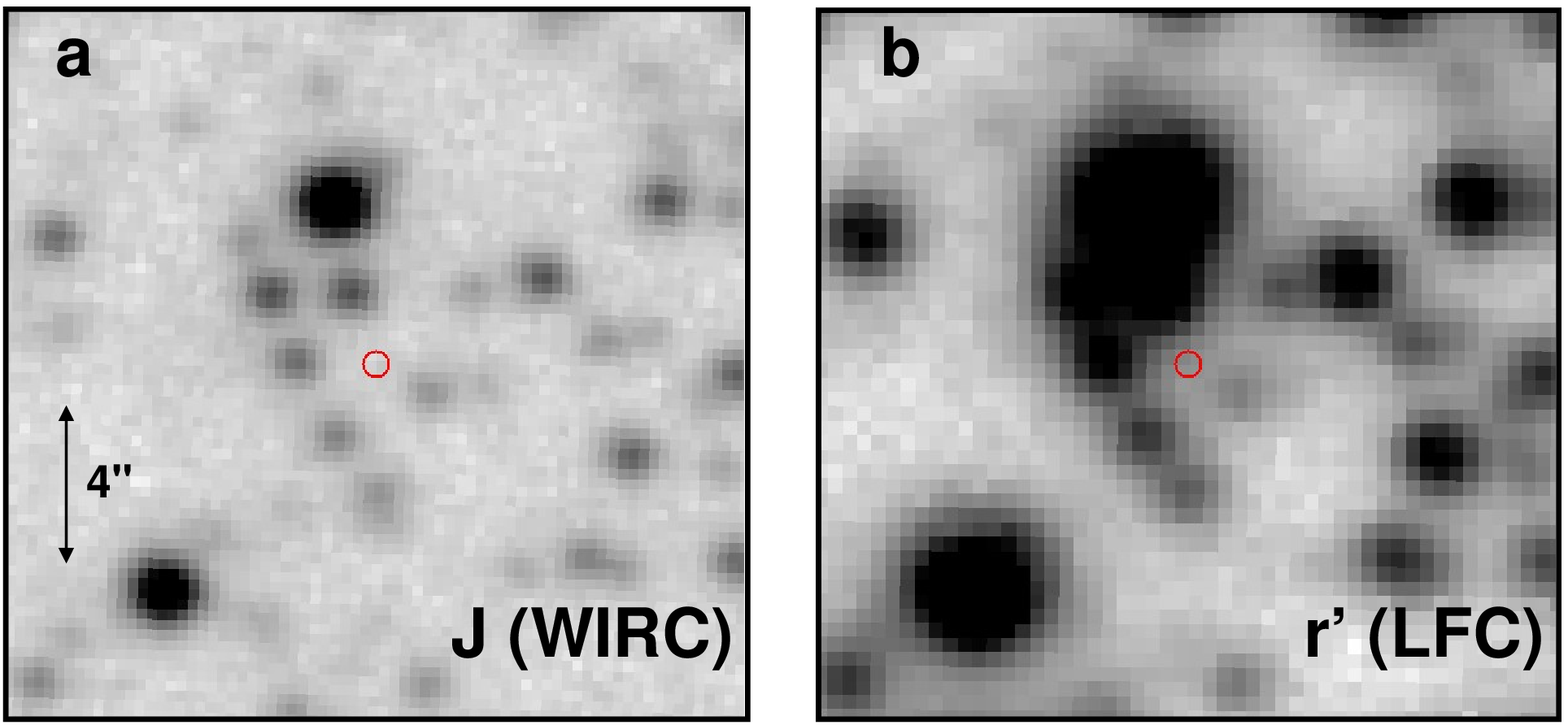 hoffset=-40 voffset=160}{14.7cm}{21.5cm}
\FigNum{\ref{fig:opt}}
\end{figure}

\clearpage

\begin{deluxetable}{lccrcr}
\tablecaption{Observations \label{tab:observations} }
\tablehead{
\colhead{Start Time } &
\colhead{} &
\colhead{Duration} &
\colhead{Source} &
\colhead{} &
\colhead{Source}\\
\colhead{(UT)} &
\colhead{Obs/Instr.} &
\colhead{(sec)} &
\colhead{Counts$^a$} &
\colhead{ECR$^b$} &
\colhead{Flux$^c$}
}
\startdata
1991 Mar 24 03:19   & {\em ROSAT}/HRI   & 4350  & $<$5      &  4.5\tee{-10} & $<$5.2  \\
1992 Oct 15 13:18  & {\em ROSAT}/PSPC  & 13861 & 10.5\ppm4  &   1.6\tee{-10} & (1.2\ppm0.4) \\
1993 Mar 24 04:40  & {\em ROSAT}/PSPC  & 12088 & 14.4\ppm4  &   \nodata          &  (1.9\ppm0.4)  \\
2001 Feb 19  11:25 & {\em Chandra}/ACIS-S  & 7787 & 97\ppm10 & 2.0\tee{-11}  &  (2.4\ppm0.2) \\
2004 Apr 13 18:30       & {\em XMM}/EPIC/pn & 8036 & $<$64  & 1.2\tee{-11} &  $<$1.0  \\
2004 Apr 19 13:36       & {\em XMM}/EPIC/pn & 21125 & $<$70  & 1.2\tee{-11} &  $<$0.4\\
2006 Mar 7   00:19 & {\em SWIFT}/XRT & 2876  & $<$4  & 9.7\tee{-11}    &  $<$1.3 \\
2006 Mar 14  00:58 & {\em SWIFT}/XRT & 2405  & $<4$  & \nodata       &   $<$1.6 \\
\enddata
\tablenotetext{a}{The expected average background number of counts
  has been subtracted, which can produce measurements of fractional
  numbers of counts. Upper limits are 90\% confidence.}

\tablenotetext{b}{Countrate to flux conversion factor, for unabsorbed
flux (\cgsflux, 0.5-10 keV) per source count/sec,
assuming \nh=0.33\tee{22} \perval{cm}{-2} and a photon power-law index
of $\alpha=0.9$. }

\tablenotetext{c}{Source Unabsorbed Flux (units of \ee{-13} \cgsflux, 0.5-10 keV)}
\end{deluxetable}

\begin{table}
\begin{center}
\caption{\label{tab:cxospec}Spectral fits to \cxo} 
\begin{tabular}{lr} \hline
\multicolumn{2}{c}{Model: Absorbed Blackbody} \\ \hline
N$_{H,22}$(cm$^{-2}$) & (0.33) \\
kT (keV) & 1.2\ud{0.3}{0.4} \\
N$_{BB}$$^a$ & 1.91\ud{1.05}{0.69}$\times10^{-6}$ \\
Model flux$^b$ & 1.43$\times10^{-13}$ \\
$\chi^2_\nu$/dof(prob) & 4.47/3(3.83$\times10^{-3}$) \\ \hline

\multicolumn{2}{c}{Model: Absorbed Powerlaw} \\ \hline
N$_{H,22}$(cm$^{-2}$) & (0.33) \\
$\alpha$$^c$ &  1.0\ud{0.3}{0.4} \\
N$_{PL}$$^d$ & 1.5\ud{0.5}{0.4}$\times10^{-5}$ \\
Model Flux $^b$& 1.8$\times10^{-13}$ \\
$\chi^2_\nu$/dof(prob) & 0.646/3(0.585) \\ \hline

\end{tabular}
\end{center}
\tablecomments{The spectral parameter fits to the \cxo \ spectrum extracted from the level=2 event file of {\it Chandra} observation 709. The extracted spectrum contained 100 counts and the background was not subtracted.  All errors are 90$\%$ confidence. \\
$^a$ L$_{39}$/D$_{10}^2$ \\
$^b$ best-fit unabsorbed flux, ergs cm$^{-2}$s$^{-1}$ (0.5-8.0keV) \\
$^c$ power-law index \\
$^d$ photons cm$^{-2}$s$^{-1}$ at 1keV }
\end{table}

\end{document}